\documentclass[aps,pre,twocolumn]{revtex4-2}

\usepackage[english]{babel}
\usepackage{amsfonts}
\usepackage{amssymb}
\usepackage{amsmath}
\usepackage{dsfont}
\usepackage{braket}
\usepackage{graphicx}
\usepackage{hyphenat}
\usepackage{hyperref}
\usepackage{placeins}	
\usepackage{xcolor}
\usepackage{csquotes}
\usepackage[nameinlink,noabbrev]{cleveref}
\usepackage{natbib}
\usepackage{changes} 
\usepackage{nth}
\usepackage{import}
\usepackage{xspace} 
\usepackage{tabularx}
\usepackage{makecell}
\usepackage{mathtools}

\setcellgapes{4pt}

\definecolor{FAU-Blau}{RGB}{0,47,108}					
\definecolor{FAU-Dunkelblau}{RGB}{4,30,66}				
\definecolor{FAU-Phil-Gelb}{RGB}{255, 184,28}			
\definecolor{FAU-Phil-Orange}{RGB}{232,119,34}			
\definecolor{FAU-RW-Rot}{RGB}{200,16,46}				
\definecolor{FAU-RW-Dunkelrot}{RGB}{151,27,47}			
\definecolor{FAU-Med-Blau}{RGB}{0,163,224}				
\definecolor{FAU-Med-Dunkelblau}{RGB}{0,97,160}			
\definecolor{FAU-Nat-Gruen}{RGB}{67,176,42}				
\definecolor{FAU-Nat-Dunkelgruen}{RGB}{34,136,72}		
\definecolor{FAU-Tech-Metallic}{RGB}{119,159,181}		
\definecolor{FAU-Tech-Dunkelmetallic}{RGB}{65,116,141}	

\hypersetup{colorlinks=true, linkcolor=FAU-Med-Dunkelblau, filecolor=FAU-Med-Dunkelblau, citecolor=FAU-Med-Dunkelblau, urlcolor=FAU-Med-Dunkelblau}

\newcommand{\identity}{\mathds{1}}
\newcommand{\ii}{\mathrm{i}}
\newcommand{\dd}{\mathrm{d}}
\DeclareMathOperator{\sgn}{sgn}

\newcommand{\lindbladian}{\mathcal{L}}
\newcommand{\operator}{\mathcal{M}}
\newcommand{\similarity}{\mathcal{S}}

\newcommand{\ketbra}[2]{\ket{#1}\!\!\bra{#2}}				
\newcommand{\bbra}[1]{\langle\!\bra{#1}}					
\newcommand{\kett}[1]{\ket{#1}\!\rangle}					
\newcommand{\bbrakett}[1]{\langle\!\braket{#1}\!\rangle}	

\newcommand{\annired}[1]{b^{\protect\vphantom{\dagger}}_{{#1},-}}
\newcommand{\creared}[1]{b^\dagger_{{#1},-}}
\newcommand{\countred}[1]{n_{{#1},-}}
\newcommand{\anniblue}[1]{b^{\protect\vphantom{\dagger}}_{{#1},+}}
\newcommand{\creablue}[1]{b^\dagger_{{#1},+}}
\newcommand{\countblue}[1]{n_{{#1},+}}
\newcommand{\annibr}[1]{b^{\protect\vphantom{\dagger}}_{{#1},\pm}}
\newcommand{\creabr}[1]{b^\dagger_{{#1},\pm}}
\newcommand{\countbr}[1]{n_{{#1},\pm}}

\newcommand{\anniminus}[1]{a^{\protect\vphantom{\dagger}}_{{#1},-}}
\newcommand{\creaminus}[1]{a^\dagger_{{#1},-}}
\newcommand{\anniplus}[1]{a^{\protect\vphantom{\dagger}}_{{#1},+}}
\newcommand{\creaplus}[1]{a^\dagger_{{#1},+}}
\newcommand{\annipm}[1]{a^{\protect\vphantom{\dagger}}_{{#1},\pm}}

\newcommand{\annialpha}[1]{a^{\protect\vphantom{\dagger}}_{{#1},\alpha}}

\newcommand{\creabeta}[1]{a^\dagger_{{#1},\beta}}
\newcommand{\annigamma}[1]{a^{\protect\vphantom{\dagger}}_{{#1},\gamma}}
\newcommand{\creagamma}[1]{a^\dagger_{{#1},\gamma}}
\newcommand{\anniboth}[1]{a^{\protect\vphantom{\ddagger}}_{{#1},\ast}}
\newcommand{\creaboth}[1]{a^\ddagger_{{#1},\ast}}
\newcommand{\anniright}[1]{a^{\protect\vphantom{\ddagger}}_{{#1},-}}
\newcommand{\crearight}[1]{a^\ddagger_{{#1},-}}
\newcommand{\annileft}[1]{a^{\protect\vphantom{\ddagger}}_{{#1},+}}
\newcommand{\crealeft}[1]{a^\ddagger_{{#1},+}}
\newcommand{\annilr}[1]{a^{\protect\vphantom{\ddagger}}_{{#1},\pm}}
\newcommand{\crealr}[1]{a^\ddagger_{{#1},\pm}}
\newcommand{\annirl}[1]{a^{\protect\vphantom{\ddagger}}_{{#1},\mp}}
\newcommand{\crearl}[1]{a^\ddagger_{{#1},\mp}}

\newcommand{\m}{\mathbf{m}}

\newcommand{\mAlphas}{{\underline{m}}}
\newcommand{\mVecAlphas}{{\underline{\mathbf{m}}}}
\newcommand*{\placeholder}{\makebox[1ex]{$\ast$}\makebox[1ex]}

\newcommand{\num}[1]{^{({#1})}}
\newcommand{\pcst}{$\textrm{pcst}^{\texttt{++}}$\xspace}
\newcommand{\nol}{q}

\definechangesauthor[name=Kai, color=FAU-Med-Blau]{KPS}
\definechangesauthor[name=Lea, color=FAU-Nat-Gruen]{LL}
\definechangesauthor[name=Andi, color=FAU-RW-Rot]{AS}

\begin{document}

\title{Series expansions in closed and open quantum many-body systems with multiple quasiparticle types}

\author{Lea Lenke}
\email{lea.lenke@fau.de}

\author{Andreas Schellenberger}
\email{andreas.schellenberger@fau.de}

\author{Kai Phillip Schmidt}
\email{kai.phillip.schmidt@fau.de}

\affiliation{Friedrich-Alexander-Universit\"at Erlangen-N\"urnberg, Department of Physics, Staudtstraße 7, 91058 Erlangen, Germany}

\begin{abstract}
	The established approach of perturbative continuous unitary transformations (pCUTs) constructs effective quantum many-body Hamiltonians as perturbative series that conserve the number of one quasiparticle type.
	We extend the pCUT method to similarity transformations -- dubbed \pcst -- allowing for multiple quasiparticle types with complex-valued energies.
	This enlarges the field of application to closed and open quantum many-body systems with unperturbed operators corresponding to arbitrary superimposed ladder spectra.
	To this end, a generalized counting operator is combined with the quasiparticle generator for open quantum systems recently introduced by Schmiedinghoff and Uhrig \cite{SchmiedinghoffUhrig2022}.
    The \pcst then yields model-independent block-diagonal effective Hamiltonians and Lindbladians allowing a linked-cluster expansion in the thermodynamic limit similar to the conventional pCUT method.
    We illustrate the application of the \pcst method by discussing representative closed, open, and non-Hermitian quantum systems.
\end{abstract}

\maketitle

\section{Introduction}

The investigation of collective behavior in correlated quantum matter is an active research field in condensed matter and quantum optics because it is relevant for the development of functional quantum materials and quantum technologies \cite{Dowling2003, Acin2018}.
Theoretically, the treatment of such quantum many-body systems is extremely challenging due to the exponential increase of the Hilbert-space dimension as a function of system size.
This is particularly true for open quantum systems described by quantum master equations because the operator-space dimension scales quadratically with the state-space dimension \cite{Banerjee2018}.
Examples are Lindblad master equations where, in contrast to Hermitian Hamiltonians and observables used for closed systems, the associated Lindblad operator is non-Hermitian, expressing the dissipative nature of the system \cite{Breuer2007, Cleve2017}.
Additionally, non-Hermitian Hamiltonians are fundamentally relevant and appear further in quantum magnetism in terms of the Dyson-Maleev representation of spin operators \cite{ElGanainy2018, Ling2022, LenkeMuehlhauserSchmidt2021}.
It is therefore an important line of research to generalize existing theoretical tools for closed Hermitian systems to open quantum many-body systems, which has been pursued in recent years, e.g., by quantum trajectories \cite{Daley2014}, tensor networks \cite{White2004, Verstraete2004, Zwolak2004, Kshetrimayum2017}, extensions of mean-field theories \cite{Jin2016, Landa2020}, and continuous similarity transformations \cite{Rosso2020, SchmiedinghoffUhrig2022}.
Of particular interest are approaches that are able to treat large or infinite systems so that physical properties can be extracted in the thermodynamic limit.

High-order series expansions are a powerful tool to investigate closed quantum many-body systems.
Typically, one exploits the linked-cluster theorem to determine the exact expression of physical quantities perturbatively up to high orders in the thermodynamic limit by performing calculations on finite linked clusters \cite{Singh1988,Gelfand1990}.
Initially, such linked-cluster expansions determined high-temperature series for extensive thermodynamic quantities as well as extensive zero-temperature ground-state properties like the ground-state energy \cite{Levy1982}.
In contrast, for non-extensive quantities like excitation energies, linked-cluster expansions are more complicated and it took until 1996 when Gelfand set up a true linked-cluster expansion for a one-particle dispersion \cite{Gelfand1996}.
However, this approach violates the cluster additivity and is therefore only applicable when the ground state and the targeted excitation subspace are characterized by different quantum numbers.
This has been resolved in 2000 with the use of orthogonal transformations on graphs so that cluster additivity is restored and linked-cluster expansions for many-particle excitation energies became possible \cite{Trebst2000, Zheng2001}.

Another attractive route to linked-cluster expansions for closed quantum many-body systems -- which fulfills cluster additivity by design -- is the method of perturbative continuous unitary transformations (pCUTs) \cite{Wegner1994, OitmaaHamerZheng2006,KnetterUhrig2000} allowing for the treatment of many-particle excitation energies as well as spectral densities \cite{KnetterSchmidtUhrig2003}. Within the pCUT approach, a quasiparticle-conserving (QP-conserving) and model-independent effective Hamiltonian in second quantization is derived, with the constraint that the unperturbed part of the Hamiltonian has an equidistant spectrum and is bounded from below \cite{KnetterUhrig2000}.
Over the last two decades pCUT was indeed applied successfully as a linked-cluster expansion, i.e., a full graph decomposition has been implemented to calculate relevant matrix elements for a large variety of correlated closed quantum many-body systems.
This includes frustrated quantum magnets \cite{Coester2013,Wagner2021}, models displaying topological quantum order \cite{Vidal2009, Muehlhauser2020, Muehlhauser2022}, and systems with long-range interactions \cite{Fey2019, Adelhardt2020} or with quenched disorder \cite{Hoermann2018, Hoermann2020}.
The latter applications exploit the presence of an effective Hamiltonian in second quantization using white-graph expansions \cite{Coester2015}.

The QP-conserving effective pCUT Hamiltonian is defined by a single QP-counting operator corresponding to the equidistant spectrum of the unperturbed Hamiltonian \cite{KnetterUhrig2000}.
It would therefore be desirable to generalize the pCUT approach to multiple QP types, while keeping the method model-independent, in contrast to other generalizations like enhanced pCUT \cite{Krull2012}.
At the same time, to describe open quantum system, a generalization of the pCUT method to non-Hermitian Hamiltonians and Lindbladians is needed.

In this work we introduce generalized perturbative continuous similarity transformations -- \pcst -- that resolve these issues.
We extend the pCUT formalism by generalizing the counting operator to multiple QP types.
For non-Hermitian operators, we additionally combine it with the generalized QP generator for open quantum systems by Schmiedinghoff and Uhrig \cite{SchmiedinghoffUhrig2022}, resulting in a similarity transformation.
As a result, model-independent Hamiltonians and Lindbladians are derived within \pcst allowing a linked-cluster expansion similar to the conventional pCUT method in the thermodynamic limit.

The paper is organized as follows.
In Sec.~\ref{sec:generalizedpcut} we describe the \pcst approach and its most important properties.
Several applications of the \pcst approach, including Hermitian and non-Hermitian Hamiltonians as well as Lindbladians, are discussed in Sec.~\ref{sec:models}.
Finally, we draw conclusions in Sec.~\ref{sec:conclusion}.

\section{Method}
\label{sec:generalizedpcut}

This section includes all technical aspects relevant for the \pcst approach.
We first introduce continuous similarity transformations (CSTs) generalizing the method of continuous unitary transformations (CUTs) \cite{Wegner1994, GlazekWilson1993}.
For the CST we use the generalized quasiparticle (QP) generator \cite{SchmiedinghoffUhrig2022}, allowing us to derive effective operators that are block diagonal.
The \pcst approach then corresponds to a model-independent perturbative solution of the underlying flow equation describing the CST.

\subsection{Continuous similarity transformations}

The goal of the CST is to map the operator describing the system, e.g., the Hamiltonian, to an effective operator that is easier to treat.
This is done in a continuous fashion.
Suppose we want to transform the operator $\operator$ into a more suitable basis.
For that we define a similarity transformation $\similarity(\ell)$ that continuously depends on a flow parameter $\ell \in [0, \infty]$.
The flowing operator
\begin{equation}
	\operator(\ell) = \similarity(\ell) \operator \similarity(\ell)^{-1}
\end{equation}
is a function of the flow parameter $\ell$ and is used to transform the operator $\operator = \operator(0)$ into the effective operator $\operator_\textrm{eff} = \operator(\infty)$.
The effective operator $\operator_\textrm{eff}$ is thus equal to $\operator$ but expressed in a more suitable basis.
Since the similarity transformation $\similarity(\ell)$ is continuous, it has an infinitesimal generator $\eta(\ell)$, i.e.,
\begin{equation}
	\partial_\ell \similarity(\ell) = - \similarity(\ell) \eta(\ell)\,. \label{eq:similarity}
\end{equation}
We can express the evolution of $\operator(\ell)$ with respect to the flow parameter $\ell$ in terms of the generator $\eta(\ell)$.
This results in the flow equation
\begin{align}
	\partial_\ell \operator(\ell) = [\eta(\ell), \operator(\ell)]\,. \label{eq:flow_equation}
\end{align}
This differential equation is typically not exactly solvable because it corresponds to an infinite number of coupled differential equations for the coefficients of operators appearing in $\operator(\ell)$.
As a consequence, one has to truncate the flow equation using appropriate truncation parameters, e.g., in a perturbative coupling \cite{KnetterUhrig2000,KnetterSchmidtUhrig2003,Krull2012}, in the scaling dimension \cite{Powalski2015, Powalski2018}, or in the spatial extension of operators \cite{Fischer2010,Drescher2010}.
The resulting finite number of differential equations can then be solved numerically for a specific class of models.

Clearly, the resulting $\operator_\textrm{eff}$ depends on the choice of the generator $\eta(\ell)$.
For Hermitian operators we use the QP generator \cite{KnetterUhrig2000, Mielke1998} that reads
\begin{equation}
	\eta(\ell)_{nm} = \sgn(Q_{nn} - Q_{mm}) \operator(\ell)_{nm} \label{eq:quasi_particle_generator}
\end{equation}
using a matrix element notation in the eigenbasis of the QP counting operator $Q$ that can be chosen model specifically.

The resulting effective operator fulfills $[Q,\operator_\textrm{eff}]=0$, thus conserving the number of QPs. 
Therefore $\operator_\textrm{eff}$ is block-diagonal, i.e., eigenstates of $Q$ corresponding to different eigenvalues are decoupled.
If the operator $Q$ is not Hermitian, we use the generalized QP generator \cite{SchmiedinghoffUhrig2022} by continuously extending the sign function as
\begin{equation}
	\sgn(z) := \begin{cases}
		0 & \textrm{for $z = 0$\,,}\\
		\frac{z^*}{|z|} = e^{-\ii \arg(z)} & \textrm{else\,.}
	\end{cases}\label{eq:complex_sgn}
\end{equation}
Like the real sign function, this extension fulfills \mbox{$z \cdot \sgn(z) = |z|$}.

\subsection{Generalized perturbative continuous similarity transformations -- \pcst }

In certain cases it is possible to perturbatively calculate the effective operator $\operator_\textrm{eff}$ independent of the specific model.
For these perturbative calculations, we need that $\operator$ can be split as
\begin{equation}
	\operator = Q + \lambda V\,,
\end{equation}
with an unperturbed part $Q$ and a perturbation $V$ with perturbation parameter $\lambda$.
The generalization to multiple perturbation parameters is straightforward.
Furthermore, we assume that
\begin{align}
	Q = \sum_{\alpha = 1}^\nol \epsilon\num{\alpha} Q\num{\alpha}\,, && \epsilon\num{\alpha}\in \mathbb C\,,
	\label{eq:Qoperator}
\end{align}
where each individual $Q\num{\alpha}$ has an equidistant ladder spectrum , which is normalized to a spacing of $1$ and bounded from below, and that the perturbation can be decomposed as
\begin{align}
	V = \sum_{m \in \mathcal{E}} T_m\,, && [Q, T_m] = m T_m\,, \label{eq:generalized_commutation_relation}
\end{align}
with $\mathcal{E}$ a finite set of complex numbers.
The index $m$ indicates the mapping from the eigenspace of $Q$ with eigenvalue $q$ to the one with $q+m$ when acting with $T_m$; it is generically not an integer number.
For $\nol = 1$ and Hermitian operators, the transformation reduces to the conventional pCUT \cite{KnetterUhrig2000}.

The most general ansatz for the operator during the flow is
\begin{equation}
\begin{split}
	\operator(\ell) &= Q + \sum_{k = 1}^{\infty} \lambda^k \sum_{|\m| = k} F(\ell; \m) T_\m\,,\\
	\m &= (m_1, \dots, m_k)\,,\\
	T_\m &= T_{m_1} \cdots T_{m_k}\,,
\end{split} \label{eq:ansatz}
\end{equation}
with $m_i \in \mathcal{E}$ and complex coefficient functions $F(\ell; \m)$.
The commutation relation from Eq.~\eqref{eq:generalized_commutation_relation} leads to
\begin{align}
	[Q, T_\m] = M(\m) T_\m\,, && M(\m) = \sum_{i = 1}^k m_i\,.
	\label{eq:commutator_Q_Tm}
\end{align}
Therefore, $M(\m)$ measures the mapping from the eigenspace of $Q$ with eigenvalue $q$ to the one with \mbox{$q+M(\m)$} when acting with $T_\m$.
The (generalized) QP generator Eq.~\eqref{eq:quasi_particle_generator} reads
\begin{equation}
	\eta(\ell) = \sum_{k = 1}^\infty \lambda^k \sum_{|\m| = k} \sgn(M(\m)) F(\ell; \m) T_\m\,. \label{eq:ansatz_generator}
\end{equation}
The flow equation \eqref{eq:flow_equation} is then equal to a set of recursive differential equations for the modified coefficient functions $f(\ell; \m) = e^{|M(\m)| \ell} F(\ell; \m)$, reading
\begin{align}
\begin{split}
	\partial_\ell f(\ell; \m) = \sum_{(\m_1, \m_2) = \m} e^{(|M(\m)| - |M(\m_1)| - |M(\m_2)|) \ell}\\
	\times [\sgn(M(\m_1)) - \sgn(M(\m_2))] f(\ell; \m_1) f(\ell; \m_2)\,,
\end{split}
\label{eq:RecursiveDifferentialEquation}
\end{align}
where the function corresponding to the empty tuple is defined as $f(\ell; ()) = 0$.
These equations can be solved order by order with starting conditions
\begin{equation}
	f(0; \m) = \delta_{1, |\m|}\,.
\end{equation}
The resulting coefficients $C_\m := F(\infty; \m)$, and thus the effective operator $\operator_\textrm{eff}$, only depend on $\mathcal{E}$ and not on the specific definitions of $Q$ or $T_m$.
That means our generalized transformation is at least as model independent as pCUT.
This is different for enhanced pCUT (epCUT) \cite{Krull2012}, which also generalizes pCUT to non-equidistant unperturbed spectra but is, however, not model independent.
It is possible to include additional degrees of freedom as variables in the prefactors $\epsilon\num{\alpha}$ and indices $m$.
Of course, this introduces further computational challenges and the maximally achievable order of perturbation is smaller than for fixed complex $\epsilon\num{\alpha}$ and $m$. Consequently, to achieve higher perturbative orders, one can always fix $\epsilon\num{\alpha}$ to numerical values to speed up solving the flow equation.

The effective operator can be written as
\begin{equation}
	\mathcal M_\mathrm{eff} = Q + \sum_{k=1}^\infty \lambda ^k \sum_{M(\m)=0} C_\m T_\m\,.
	\label{eq:Meff}
\end{equation}
Due to the restriction of the sum to \mbox{$M(\m)=0$}, $\operator_\mathrm{eff}$ fulfills \mbox{$[Q, \operator_\textrm{eff}] = 0$} and is thus block diagonal.
The proof for the band diagonality during the flow, the block diagonality of the effective operator, and the convergence of the \pcst follows closely the one of pCUT \cite{KnetterUhrig2000} and is explained in the \hyperref[sec:proof_trafo]{Appendix}.

As \pcst generalizes pCUT, it also realizes a linked-cluster expansion, as discussed for pCUT in \cite{Gelfand1990,KnetterSchmidtUhrig2003, PhdKnetter2003}.
This enables us to determine physical quantities in the thermodynamic limit by only using finite clusters.
To obtain physical quantities, we calculate matrix elements of the form $\braket{s_2 | \operator_\mathrm{eff} | s_1}$ with model-specific states $\ket{s_1}$, $\ket{s_2}$ on sufficiently large finite clusters.
For computation, we can rely on the same programs used for pCUT evaluations, explained in more detail in \cite{KnetterUhrig2000}.
Therefore, we are able to determine physical quantities on the same kind of lattices and interactions equally efficient as for pCUT (e.g.\ long-range interaction \cite{Coester2015, Fey2019}, 2D systems \cite{Dorier2008, Vidal2009}, 3D systems \cite{Muehlhauser2020,Muehlhauser2022}) in the thermodynamic limit.
In summary after the bare transformation the subsequent linked-cluster expansion and thus the computational difference with respect to different interactions or dimensions is equal for pCUT as for \pcst.

For $\operator_\mathrm{eff}$, it is easily possible to calculate the coefficient functions by hand, up to second order.
The resulting coefficients read
\begin{equation}
\begin{split}
	C_{(m_1)} &= \begin{cases}
		1 & \textrm{for $M((m_1)) = 0$\,,}\\
		0 & \textrm{else\,,}
	\end{cases}\\
	C_{(m_1, m_2)} &= \begin{cases}
		\frac{1}{M((m_1))} & \textrm{for $M((m_1)) = - M((m_2)) \neq 0$\,,}\\
		0 & \textrm{else\,.}
	\end{cases}
\end{split} \label{eq:coefficientsO1O2}
\end{equation}
The number of indices $\m$ and the complexity of the calculations increase drastically with increasing order.
Therefore, it is useful to automate the solving of the flow equation in a similar fashion as for the conventional pCUT method \cite{KnetterUhrig2000}.
The program we used for calculating the transformation is freely accessable on GitHub~\cite{Program}.

If in addition observables, states, or eigenvectors are required in the effective basis, we can use \pcst to calculate the similarity transformation $\similarity(\infty)$ explicitly by inserting Eq.~\eqref{eq:ansatz_generator} and the ansatz
\begin{equation}
	\similarity(\ell) = \identity + \sum_{k = 1}^\infty \lambda^k \sum_{|\m| = k} G(\ell; \m) T_\m
\end{equation}
into Eq.~\eqref{eq:similarity} and solving for $F(\ell; \m)$ and $G(\ell; \m)$ recursively.
Alternatively, it is possible to make a similar ansatz explicitly for a given observable.

\subsection{Coupling of subspaces with small energy spacing}
\label{sec:broad_signum}
As a perturbative transformation, the series obtained by \pcst are limited by a certain convergence radius, depending qualitatively on the ratio between the unperturbed energy levels and the perturbation $\lambda$.
For the conventional pCUT method, the unperturbed Hamiltonian is limited to a single ladder spectrum, with a fixed spacing that can be normalized to 1.
Thus for $\lambda \ll 1$, the perturbative approach is justified.
In contrast, the more general structure of $Q$ in Eq.~\eqref{eq:Qoperator} for \pcst allows for more versatile energy spacings in the unperturbed Hamiltonian depending on the ratio of the $\epsilon\num{\alpha}$, including arbitrary small spacings.
If virtual processes with these unperturbed energy spacings occur, the convergence radius is limited by them.
Exemplary, this can be seen in the second-order coefficient $C_{(m_1,m_2)}$ in Eq.~\eqref{eq:coefficientsO1O2}, which can be the inverse of the unperturbed energy difference.

To increase the radius of convergence for systems with small energy spacings in $Q$, we adjust the generator in Eq.~\eqref{eq:quasi_particle_generator} by introducing the broad-step signum function
\begin{equation}
	\sgn_{D}(x) := \begin{cases}
		0 & \text{if } |x| \leq D\,,\\
		\sgn (x) & \text{else}\,,
	\end{cases}
    \label{eq:broad_signum}
\end{equation}
to prevent the \pcst from decoupling subspaces with an unperturbed energy spacing not more than $D$.
The effective Hamiltonian can then include processes where the unperturbed energy changes up to $D$.
Thus it does not necessarily commute with $Q$ and larger blocks are formed, in general.
Blocks that are decoupled by the original \pcst might now require additional post-diagonalization.
We will exemplary show the increased convergence in Sec.~\ref{sec:commuting}.

\subsection{Special case of multiple quasiparticle types}
\label{sec:multipleqptypes}

Assume the special case of $\nol$ different QP types, each counted by an individual number operator $Q\num{\alpha}$.
Further assume that the perturbation additionally fulfills \mbox{$[Q\num{\alpha}, T_m] = m\num{\alpha} T_m$} with an integer $m\num{\alpha}$ for all ${\alpha \in \{1,\dots,\nol\}}$.
In that case, we can replace the label $m$ by the redundant label ${\mAlphas = (m\num{1}, \dots, m\num{\nol})}$ with ${m = \sum_{\alpha = 1}^\nol \epsilon\num{\alpha} m\num{\alpha}}$ to get a more intuitive picture of the perturbation operators.
Physically, the index $m\num{\alpha}$ indicates how many QPs of type $\alpha$ are created or annihilated by $T_\mAlphas$.
In this notation, we can follow the same ansatz as done in Eq.~\eqref{eq:ansatz} to obtain an effective operator analog to Eq.~\eqref{eq:Meff} as
\begin{equation}
	\mathcal M_\mathrm{eff} = Q + \sum_{k=1}^\infty \lambda ^k \sum_{M(\mVecAlphas)=0} C_{\mVecAlphas} T_\mVecAlphas\,. \label{eq:Meff_special}
\end{equation}
Hereby, we define the vectors $\mVecAlphas = (\mAlphas_1, \dots, \mAlphas_k)$, the operators $T_{\mVecAlphas} = T_{\mAlphas_1} \cdots T_{\mAlphas_k}$, and the function
\begin{equation}
	M(\mVecAlphas) =\sum_{\alpha=1}^\nol \epsilon\num{\alpha} \sum_{i = 1}^k m\num{\alpha}_i
\end{equation}
fulfilling $[Q, T_\mVecAlphas] = M(\mVecAlphas) T_\mVecAlphas$.
As for Eq.~\eqref{eq:commutator_Q_Tm}, this function can be understood as the mapping from the eigenspace of $Q$ with eigenvalue $q$ to the one with \mbox{$q+M(\mVecAlphas)$} when acting with $T_\mVecAlphas$.
In the same way, the relation ${[Q\num{\alpha}, T_\mVecAlphas] = \sum_{i = 1}^k m\num{\alpha}_i}$ describes how many QPs of type $\alpha$ are created upon acting with $T_\mVecAlphas$.

The effective operator is again block diagonal because of the vanishing commutator $[Q, \operator_\textrm{eff}] = 0$ due to $M(\mVecAlphas) = 0$ in Eq.~\eqref{eq:Meff_special}.
In contrast, the individual $[Q\num{\alpha}, \operator_\textrm{eff}]$ may be non-zero, because $M(\mVecAlphas) = 0$ does not imply $\sum_{i = 1}^k m\num{\alpha}_i = 0$.
This gives rise to potential QP-type conversion, depending on the model, as we will discuss exemplary in Sec.~\ref{sec:models}.

Due to the Jacobi identity
\begin{equation}
	[Q\num{\alpha}, [Q\num{\beta}, \placeholder]] - [Q\num{\beta}, [Q\num{\alpha}, \placeholder]] = [[Q\num{\alpha}, Q\num{\beta}], \placeholder]\,,
\end{equation}
we can use the notation in terms of multiple QP types if and only if $[[Q\num{\alpha}, Q\num{\beta}], V] = 0$ for all $\alpha, \beta \in \{1, \dots, \nol\}$.
This implies that $V$ is in the subspace where $[Q\num{\alpha}, \placeholder]$ and $[Q\num{\beta}, \placeholder]$ commute, which, in turn, enables us to solve the eigenequations $[Q\num{\alpha}, T_\mAlphas] = m\num{\alpha} T_\mAlphas$ for all \mbox{$\alpha \in \{1,\dots,\nol\}$} simultaneously.
In this eigenequation, the operator $T_\mAlphas$ is an eigenvector of the operator $[Q\num{\alpha}, \placeholder]$ corresponding to the eigenvalue $m\num{\alpha}$.

Whenever possible we use this notation, as it closer resembles the original pCUT method.

\section{Applications}
\label{sec:models}
After introducing the \pcst, this section aims at showing the versatile models that can be treated with the generalized transformation.
This includes models with different types of QPs in separate and common Hilbert spaces (see Secs.~\ref{sec:commuting} and \ref{sec:non-commuting}, respectively), non-Hermitian Hamiltonians (see Sec.~\ref{sec:non-hermitian}), and open systems described by a Lindbladian rather than a Hamiltonian (see Sec.~\ref{sec:lindbladian}).
We keep the physical discussion on the single models rather short to focus on the technical aspects of \pcst and its capabilities. As the number of terms in Eq.~\eqref{eq:Meff} grows very fast in perturbation order, we used a computer program to calculate the action of the effective operators on specific states.

\subsection{Staggered transverse-field Ising model}
\label{sec:commuting}

\begin{figure}
    \centering
    \includegraphics[width=\columnwidth]{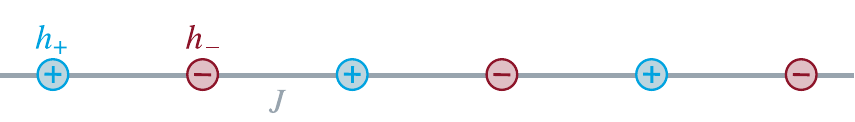}
    \caption{
    Transverse field Ising chain in a staggered magnetic field. The magnetic field has alternating strengths $h_+$ and $h_-$ indicated by blue and red circles, respectively. Nearest-neighbor spins are coupled by an Ising interaction of strength $J$, marked in gray.}
    \label{fig:chain}
\end{figure}

As a first model, we consider the transverse-field Ising chain with a staggered magnetic field (STFIM) about the high-field limit, with two alternating field strengths $h_\pm = h \pm \delta_h/2$, with $h, \delta_h \in \mathbb R$, given by the Hamiltonian
\begin{align}
    \begin{split}
        \mathcal H_\mathrm{STFIM}= h_+ \sum_{j \in +} \sigma^z_j  +   h_- \sum_{j \in -} \sigma^z_j - J \sum_{\braket {i,j}} \sigma^x_i \sigma^x_j\, .
    \end{split}
\end{align}
Without loss of generality, we consider the case ${h> \delta_h \geq 0}$ and ferromagnetic Ising interactions $J\geq 0$.
In the high-field limit $h_\pm > J$, one has two types of QPs on the alternating sets of sites denoted by $\pm$ (see Fig.~\ref{fig:chain}).
As the different QP types exist only on separate sites, the QPs live in distinct Hilbert spaces and only couple due to the perturbation induced by the Ising interaction.

By applying the Matsubara-Matsuda transformation \cite{Matsubara1956} we can write $\mathcal H_\mathrm{STFIM}$ in terms of hardcore-bosonic creation and annihilation operators of two QP types:
\begin{align}
    \begin{split}
        \mathcal H_\mathrm{STFIM}=&\, E_0 + 2h_+ Q\num{1} + 2h_- Q\num{2} \\
		&- J \left[ T_{(+1,+1)} +T_{(+1,-1)} + \mathrm{h.c.}\right]\,,
    \end{split}
    \label{eq:HSTFIM_2quant}
\end{align}
with the bare ground-state energy ${E_0 = -Nh}$, $N$ the even number of sites, the number operators
\begin{align}
	Q\num{1} &= \sum_{j \in +} \creablue{j} \anniblue{j}, & Q\num{2} &= \sum_{j \in -} \creared{j} \annired{j}
    \label{eq:HSTFIM_Q}
\end{align}
for the two QP types, and the $T$-operators given in Tab.~\ref{tab:hamiltonian-commuting}.
The number operators commute, so we can split the index of the perturbation operators $T_{(m_1,m_2)}$ with respect to the two QP-types, as discussed in Sec.~\ref{sec:multipleqptypes}.
The energy quanta from the individual ladder spectra from Eq.~\eqref{eq:HSTFIM_Q} are given by \mbox{$\epsilon\num{1} = 2h_+$} and $\epsilon\num{2}= 2h_-$.

{\makegapedcells\begin{table}
	\begin{tabularx}{\columnwidth}{c|c|l}
		$m\num{1}$ & $m\num{2}$ & Terms in $T_\mAlphas$\\ \hline\hline
		$+1$ & $+1$ & $\creablue{i} \creared{j}$\\ \hline
		$+1$ & $-1$ & $\creablue{i} \annired{j}$\\ \hline
		$-1$ & $+1$ & $\anniblue{i} \creared{j}$\\ \hline
		$-1$ & $-1$ & $\anniblue{i} \annired{j}$
	\end{tabularx}
	\caption{Perturbation operators of the STFIM Eq.~\eqref{eq:HSTFIM_2quant}. The $T$-operators fulfill the relation  ${T_{(m\num{1}, m\num{2})}^\dagger = T^{\phantom{\dagger}}_{(-m\num{1}, -m\num{2})}}$, as the model is Hermitian.\label{tab:hamiltonian-commuting}}
\end{table}}

The case $\delta_h = 0$ can be solved by conventional pCUT because the unperturbed Hamiltonian is equidistant.
We therefore focus on the case $\delta_h > 0$ and use \pcst to calculate the ground-state energy and the excitation energies of the two 1QP sectors.
As the model is translational invariant we can calculate the 1QP dispersion relations perturbatively exact in momentum space in the thermodynamic limit, using Fourier transformation and the property that \pcst is a linked-cluster expansion \cite{LenkeMuehlhauserSchmidt2021}.

The resulting excitation gaps $\Delta_\pm$ for both QP types at zero momentum and $\delta_h > 0$ up to order 6 are given by
\begin{equation}
    \Delta_\pm = 2h_\pm \pm 2\frac{J^2}{\delta_h} \mp 2\frac{J^4}{\delta_h^3} \pm 4\frac{J^6}{\delta_h^5}\,.
    \label{eq:gap_stfim}
\end{equation}
The perturbation scales with inverse powers of the unperturbed energy difference $\delta_h$ (see Eq.~\eqref{eq:coefficientsO1O2} for comparison), setting the scale of convergence.
For $\delta_h \to 0$, this leads to arbitrary small convergence radii in $J$, due to the divergence of $\Delta_\pm$ at $\delta_h = 0$.

To increase the convergence radius, we make use of the broad-step signum function $\sgn_D$ introduced in Sec.~\ref{sec:broad_signum}.
As the two 1QP channels are not decoupled anymore by \pcst, we post-diagonalize the combined $2\times 2$ block in the thermodynamic limit in momentum space.

\begin{figure}
    \centering
    \includegraphics[width=\columnwidth]{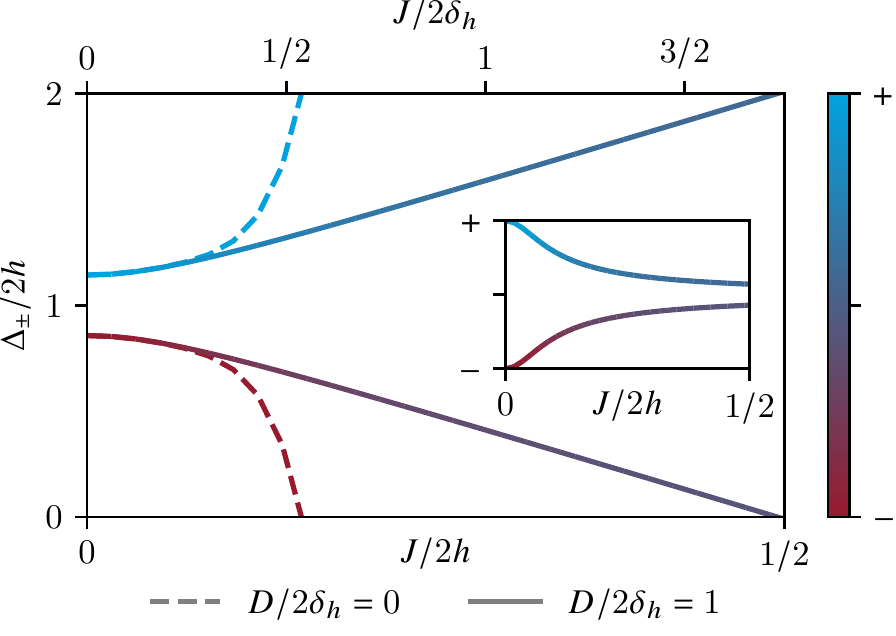}
    \caption{Normalized gaps $\Delta_\pm /2h$ of the two 1QP channels as a function of $J/2h$ and $J/2\delta_h$. The unperturbed energies are set to $h/\delta_h = 7/2$. The gaps calculated with the original \pcst are shown as dashed lines and the ones with the broad-step signum function, coupling the two 1QP channels, are shown as solid lines by setting $D = 2\delta_h = 2h_+-2h_-$. For the coupled channels, the hybridization between the two types is given in the inset plot and a color code, with blue (red) being a quasiparticle of type 1 (2) sitting on the $+$ ($-$) sites. The inset plot shows the proportion of the two QP types ($\pm$) for the two modes, using the same color code. Thereby, the limit $+$ ($-$) indicates a pure quasiparticle of type 1 (2).}
    \label{fig:spectrum_stfim}
\end{figure}

We plot the gap of the two 1QP channels in Fig.~\ref{fig:spectrum_stfim} for a fixed ratio of $h/\delta_h = 7/2$, once with the original \pcst (in dashed lines) and once with the adjusted generator (in solid lines) with $D = 2\delta_h$ to couple the subspaces of the two 1QP sectors.
The hybridization between the two QP types is given in color code and in the inset plot to visualize the enlargement of the hybridization for increasing perturbation.
As can be seen, the divergence of the original \pcst directly corresponds to a increasing hybridization of the two QP types, which is suppressed in the original method.
Using the broad-step signum function therefore enlarges the radius of convergence that is otherwise limited by the unperturbed energy difference $2\delta_h$.

\subsection{Non-Hermitian staggered transverse-field Ising model}
\label{sec:non-hermitian}

With the help of the generalized QP generator \cite{SchmiedinghoffUhrig2022}, it is possible to treat non-Hermitian Hamiltonians.
As an example, we consider the same model as in the previous section but with a purely imaginary $\delta_h \in \ii \mathbb{R}$, i.e., a non-Hermitian staggered field.
This is equal to the model considered in \cite{LiZhangZhangSong2014, LenkeMuehlhauserSchmidt2021} up to an irrelevant constant.
To increase comparability to these works, we chose the parameters $h = \eta$ and $\delta_h = 2 \ii \xi$.
In contrast to pCUT, with \pcst it is possible to treat this problem in the high-field limit $|\eta \pm \ii \xi| > |J|$.
We solve for the ground-state energy per site
\begin{equation}
	e_0 = - \frac{J^2}{4 \eta} - \frac{J^4}{64 \eta^3} \frac{\eta^2 - 3 \xi^2}{\eta^2 + \xi^2} - \frac{\eta^2 + 5 \xi^2}{\eta^2 + \xi^2} \frac{J^6}{1024 \eta^5}
\end{equation}
and the energy gaps $\Delta_\pm$ of the elementary excitations at zero momentum up to \nth{6} order
\begin{equation}
	\Delta_\pm = 2 (\eta \pm \ii \xi) \pm \frac{J^2}{\ii \xi} \pm \frac{J^4}{4 \ii \xi^3} \pm \frac{J^6}{8 \ii \xi^5}\,,
\end{equation}
being the same result as Eq.~\eqref{eq:gap_stfim}, expressed with the new variables.
We exactly reproduce the results from our previous work in \cite{LenkeMuehlhauserSchmidt2021} now using the \pcst approach instead of Takahashi perturbation theory \cite{Takahashi_1977}.
While with the latter an additional diagonalization of the resulting $2 \times 2$ block after Fourier transformation is needed, \pcst directly diagonalizes the two bands by design.

\subsection{Spin-one transverse-field Ising model with single-ion anisotropy}
\label{sec:non-commuting}
\begin{figure}
    \centering
    \includegraphics[width=\columnwidth]{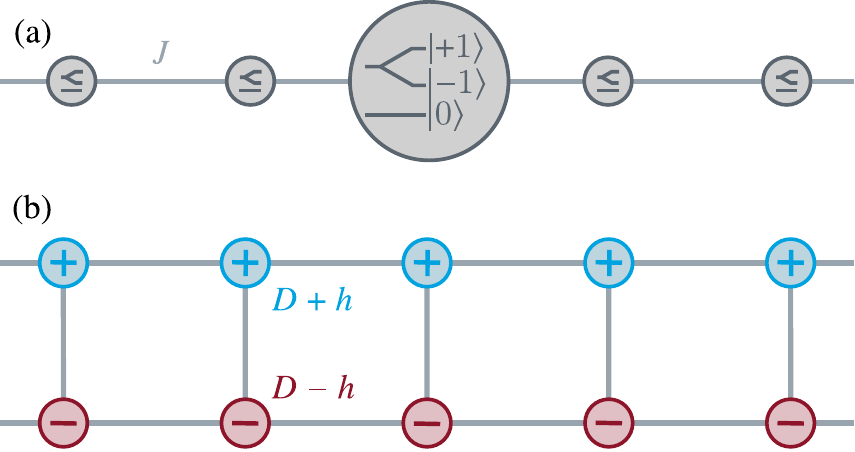}
    \caption{(a) ATFIM having local states $\ket{0}$, $\ket{+1}$, $\ket{-1}$ with unperturbed energies $0$, $D+h$, $D-h$, respectively, and perturbative Ising couplings of strength $J$.
    (b) Effective spin-$1/2$ model on a ladder according to Eq.~\eqref{eq:hamiltonian-non-commuting-transf}. The excited states $\ket{+1}, \ket{-1}$ from the spin-1 model in (a) are mapped to hardcore bosons realizing two different QP types, colored in blue and red, respectively. The Ising couplings are realized by effective four-particle interactions, as given in Tab.~\ref{tab:hamiltonian-non-commuting}.
    }
    \label{fig:spin1}
\end{figure}

The different QP types are not restricted to separate Hilbert-spaces but can also share one.
This enables us to study models with a more complicated local structure of the unperturbed Hamiltonian.
As an example we investigate the spin-1 transverse-field Ising model with single-ion anisotropy (ATFIM)
\begin{equation}
    \mathcal H_\mathrm{ATFIM} = \sum_j  \left[D\left(S_j^z\right)^2 + h S_j^z\right] - J \sum_{\braket{i,j}} S_i^x S_{j}^x\,,
\end{equation}
which is illustrated in Fig.~\hyperref[fig:spin1]{\ref*{fig:spin1}(a)}.
In the following we consider the Ising interaction as a perturbation.
By tuning the real parameters $D,h \in \mathbb R$, we can model any local three-level system with arbitrary energy spacings for the unperturbed Hamiltonian $\mathcal H_\mathrm{ATFIM} (J=0)$.
We use the eigenstates of the operators $S^z$ as our basis, i.e., locally \mbox{$S^z \ket{0} = 0$} and $S^z \ket{\pm 1} = \pm \ket{\pm 1}$.
In this basis,
\begin{equation}
	S_j^x = \frac{1}{\sqrt{2}} \big(\ketbra{0}{+1}_j + \ketbra{0}{-1}_j + \textrm{h.c.}\big)\,.
\end{equation}
For the discussion, we restrict ourselves to the parameter regime $D>h>0$, the other parameter-subspaces can be treated analogously.
Within this regime, the unperturbed local ground state is $\ket{0}$ and the two excited states are $\ket{\pm 1}$.
We define their annihilation operators as
\begin{align}
    \annipm{j} = \ketbra{0}{\pm 1}_j
\end{align}
and the respective creation operators as their adjoint operators.
We write the counting operators of the two QP types as
\begin{align}
    Q\num{1} &= \sum_{j} \creaplus{j} \anniplus{j}, & Q\num{2} &= \sum_{j} \creaminus{j} \anniminus{j}\,.
\end{align}
In contrast to Sec.~\ref{sec:commuting}, the creation- and annihilation-operators of different QP types at the same site do not commute
\begin{equation}
    [\annialpha{j}, \creabeta{j}] = \delta_{\alpha, \beta} - \creabeta{j} \annialpha{j} - \delta_{\alpha, \beta} \sum_\gamma \creagamma{j} \annigamma{j}\,.
    \label{eq:commutatorSpin1}
\end{equation}
Instead, they fulfill mutual hardcore-bosonic statistics.
Regardless of that, the operators $Q\num{1}$ and $Q\num{2}$ commute so we can split the index of the perturbation operators $T_{(m_1,m_2)}$ with respect to the two QP types, as discussed in Sec.~\ref{sec:multipleqptypes}.

We again use the translational invariance of the model to calculate the 1QP dispersions perturbatively in momentum space.
The resulting excitation gaps $\Delta_\pm$ for $D>h>0$ and $J>0$ at momentum $k=0$ up to order 4 are given by
\begin{align}
    &\begin{aligned}
        \Delta_\pm&|_{D\neq \pm 2h} = 
            D\pm h-J \pm J^{2}\frac{2 D^{2} - h^{2}}{4Dh(D \mp h)} \\ 
            & \begin{aligned}
                \pm J^3\frac{8 D^{4} \mp 8 D^{3} h \pm D h^{3} + h^{4}}{16 D^{2} h \left(D \mp h\right)^{2} \left(D \pm h\right)}
            \end{aligned} \\ 
            & \begin{aligned}
                \mp J^4 &\big[8 D^{9} \mp 24 D^{8} h \pm 32 D^{6} h^{3} + 34 D^{5} h^{4}\\ 
                &\phantom{\big[} \mp 72 D^{4} h^5 + 27 D^{3} h^{6} - 7 D h^{8} \pm 6 h^{9}\big]\\
                \times &\big[64 D^{3} h^{3} \left(D \mp h\right)^{3} \left(D \pm h\right)^{2}\left(D \mp 2 h\right)\big]^{-1} 
            \end{aligned}
        \end{aligned}\\
    &\begin{aligned}
        \left. \Delta_+\right|_{D= 2h} = \;3h - J + \frac{7J^{2}}{8 h} + \frac{67J^{3}}{192 h^{2}} -  \frac{133J^{4}}{4608 h^{3}}\,.
    \end{aligned}
\end{align}
Although, we have calculated the general expression up to order 6, the gaps are given only up to lower order, to keep the expressions compact. Higher orders for fixed $D,h$ ratios can be calculated easily, while the general analytical result, as shown here, gets hard to calculate for higher orders.
Again the perturbation scales with powers of unperturbed energy differences, corresponding to virtual processes in the perturbation theory.
Therefore, for special parameter configurations of $D,h$, like $D=2h$ starting in 4th order, we again obtain QP conversions surrounded by artificial divergences, which we can solve with the broad-step signum function in Eq.~\eqref{eq:broad_signum}.

Instead of working with the spin-1 model directly, we can also establish a mapping from the spin-1 model on a chain to two effective spin-$1/2$ degrees of freedom distributed on a ladder, as visualized in Fig.~\hyperref[fig:spin1]{\ref*{fig:spin1}(b)}, where each rung corresponds to the two QP types on one site.
Therefore, the QP types are again defined on distinct Hilbert spaces.
This effectively reduces the model to the class of systems discussed in Sec.~\ref{sec:commuting}.
Each spin-$1/2$ degree of freedom is denoted as $\ket{\uparrow}$, $\ket{\downarrow}$ and the local states on each rung are given by $\ket{\downarrow \downarrow} := \ket{0}$ for the local ground state and as $\ket{\uparrow \downarrow} := \ket{+1}$ and $\ket{\downarrow \uparrow} := \ket{-1}$ for the two corresponding excited states.
To restrict the local four-dimensional Hilbert space to the three-dimensional subspace of the spin-$1$, we use projection operators.
The Hamiltonian reads
\begin{equation}
    \begin{split}
        \mathcal H_\mathrm{ATFIM} =&\, (D+h) Q\num{1} + (D-h)Q\num{2}\\
        &- \frac{J}{2} \left(T_{(2,0)} + T_{(0,2)} + T_{(1,1)} + T_{(1,-1)} + \mathrm{h.c.}\right)
    \end{split}
    \label{eq:hamiltonian-non-commuting-transf}
\end{equation}
with $Q\num{1} = \sum_j \countblue{j}$, $Q\num{2} = \sum_j \countred{j}$ -- using the number operators ${\countbr{j} = \creabr{j}\annibr{j}}$ -- and the $T$-operators given in Tab.~\ref{tab:hamiltonian-non-commuting}, with $\annibr{j}$ ($\creabr{j}$) being the annihilation (creation) operator of the respective spin-$1/2$ on rung $j$.
Note that the two-site interactions of the Ising coupling are modeled as effective four-site interactions due to the projectors $(1-\countbr{j})$, to stay within the local three-dimensional subspace.

{\makegapedcells\begin{table}
	\begin{tabularx}{\columnwidth}{c|c|l}
		$m\num{1}$ & $m\num{2}$ & Terms in $T_\mAlphas$\\ \hline\hline
		$2$ & $0$ & $\creablue{i} \creablue{j} (1-\countred{i}) (1-\countred{j})$\\ \hline
		$0$ & $2$ & $\creared{i} \creared{j} (1-\countblue{i}) (1-\countblue{j})$\\ \hline
		$1$ & $1$ & \makecell{$\creablue{i} \creared{j} (1-\countred{i}) (1-\countblue{j})$\\ $+\, \creared{i} \creablue{j} (1-\countblue{i}) (1-\countred{j})$}\\ \hline
		$1$ & $-1$ & $\creablue{i} \annired{j} (1-\countred{i}) + \annired{i} \creablue{j} (1-\countred{j})$\\ \hline
        $0$ & $0$ & $\creablue{i} \anniblue{j} (1-\countred{i})+ \creared{i} \annired{j} (1-\countblue{i})$
	\end{tabularx}
	\caption{Perturbation operators of the ATFIM \eqref{eq:hamiltonian-non-commuting-transf}. The omitted $T$-operators are given by the Hermitian conjugate of the above expressions.\label{tab:hamiltonian-non-commuting}}
\end{table}}

To summarize, this example opens up a large class of systems with arbitrary local spectra, previously not accessible by pCUT.
By mapping the single local excitations to different QP types with mutual hardcore-bosonic statistics, as given in Eq.~\eqref{eq:commutatorSpin1}, we can establish a general procedure to calculate series expansions around these local spectra.

\subsection{Dissipative transverse-field Ising model}
\label{sec:lindbladian}

\begin{figure}
    \centering
    \includegraphics[width=\columnwidth]{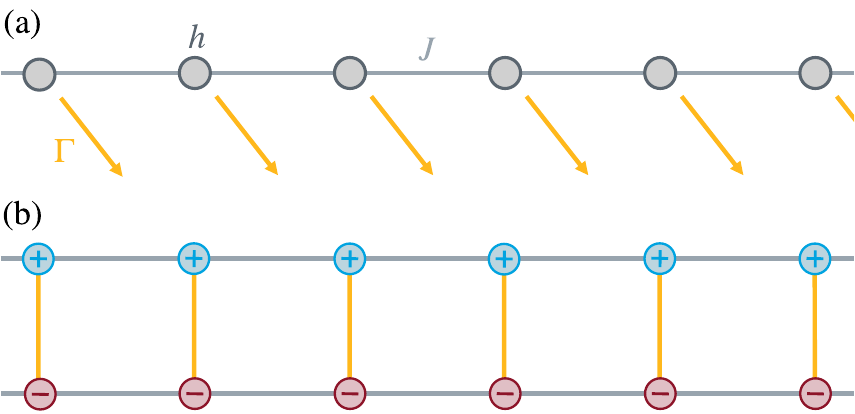}
	\caption{(a) Dissipative transverse-field Ising model. The neighboring spin-1/2 particles are coupled with an Ising interaction of strength $J$ and subject to a transverse magnetic field with strength $h$ and local dissipation with rate $\Gamma$.
	(b) Equivalent model on a ladder according to the purified Lindbladian from Eq.~\eqref{eq:Lindbladian}. Blue ($+$) sites correspond to $\sigma_j \otimes \identity$ and red ($-$) sites to $\identity \otimes \sigma_j$.
	}
	\label{fig:lindbladian}
\end{figure}

To show how versatile \pcst is, we consider the Lindbladian of the transverse field Ising model with local decay.
The coherent part of the dynamics is governed by the Hamiltonian
\begin{equation}
	\mathcal{H} = h \sum_j \sigma^z_j - J \sum_{\braket{i,j}} \sigma^x_i \sigma^x_j
\end{equation}
and the dissipative part by the local jump operators
\begin{equation}
	L^{\hphantom{-}}_j = \sigma^-_j\,,
\end{equation}
acting on each spin $j$ separately, with dissipation rate $\Gamma$.
Both parts are illustrated in Fig.~\hyperref[fig:lindbladian]{\ref*{fig:lindbladian}(a)}.
{\makegapedcells\begin{table}[t]
	\begin{tabularx}{\columnwidth}{c|c|l}
		$m\num{1}$ & $m\num{2}$ & Terms in $T_\mAlphas$\\ \hline\hline
		$- 2$ & $- 2$ & $\sigma^-_i \sigma^-_j \otimes \identity - \sigma^-_i \sigma^-_j \otimes \sigma^z_i \sigma^z_j$\\ \hline
		$+ 2$ & $- 2$ & $- \identity \otimes \sigma^-_i \sigma^-_j + \sigma^z_i \sigma^z_j \otimes \sigma^-_i \sigma^-_j$\\ \hline
		$+ 2$ & $+ 2$ & \makecell{$\sigma^+_i \sigma^+_j \otimes \identity + \sigma^z_i \sigma^+_j \otimes \sigma^-_i + \sigma^+_i \sigma^z_j \otimes \sigma^-_j$\\
		$+ \sigma^z_i \sigma^z_j \otimes \sigma^-_i \sigma^-_j$}\\ \hline
		$- 2$ & $+ 2$ & \makecell{$- \identity \otimes \sigma^+_i \sigma^+_j - \sigma^-_i \otimes \sigma^z_i \sigma^+_j - \sigma^-_j \otimes \sigma^+_i \sigma^z_j$\\
		$- \sigma^-_i \sigma^-_j \otimes \sigma^z_i \sigma^z_j$}\\ \hline
		$0$ & $0$ & \makecell{$\sigma^+_i \sigma^-_j \otimes \identity + \sigma^-_i \sigma^+_j \otimes \identity + \sigma^z_i \sigma^-_j \otimes \sigma^-_i$\\
		$+ \sigma^-_i \sigma^z_j \otimes \sigma^-_j - \identity \otimes \sigma^+_i \sigma^-_j - \identity \otimes \sigma^-_i \sigma^+_j$\\
		$- \sigma^-_i \otimes \sigma^z_i \sigma^-_j - \sigma^-_j \otimes \sigma^-_i \sigma^z_j$}\\ \hline
		$0$ & $- 2$ & \makecell{$- \sigma^z_i \sigma^-_j \otimes \sigma^-_i - \sigma^-_i \sigma^z_j \otimes \sigma^-_j + \sigma^-_i \otimes \sigma^z_i \sigma^-_j$\\
		$+ \sigma^-_j \otimes \sigma^-_i \sigma^z_j$}\\ \hline
		$+ 2$ & $0$ & $- \sigma^z_i \sigma^+_j \otimes \sigma^-_i - \sigma^+_i \sigma^z_j \otimes \sigma^-_j - 2 \sigma^z_i \sigma^z_j \otimes \sigma^-_i \sigma^-_j$\\ \hline
		$- 2$ & $0$ & $\sigma^-_i \otimes \sigma^z_i \sigma^+_j + \sigma^-_j \otimes \sigma^+_i \sigma^z_j + 2 \sigma^-_i \sigma^-_j \otimes \sigma^z_i \sigma^z_j$
	\end{tabularx}
	\caption{Perturbation operators of the Lindbladian \eqref{eq:Lindbladian}. \label{tab:Lindbladian}}
\end{table}}
Combined, the time evolution of the density matrix is given by the Lindblad equation
\begin{equation}
	\ii \dot{\rho} = [\mathcal{H}, \rho] + \frac{\ii \Gamma}{2} \sum_j \left[2 L_j^\dagger \rho L^{\hphantom{\dagger}}_j - \{L_j^\dagger L^{\hphantom{\dagger}}_j, \rho\}\right].
\end{equation}
To purify this equation, we represent the basis vectors $\ketbra{\varphi}{\psi}$ in operator space by basis vectors \mbox{$\kett{\varphi, \psi} := \ket{\varphi} \otimes \ket{\psi}^\ast$} in Liouville space.
The purification of an operator $\operator$ is defined as \mbox{$\kett{\operator} := \sum_{\ket{\varphi}, \ket{\psi}} \bra{\varphi} \operator \ket{\psi} \kett{\varphi, \psi}$}.
The Lindbladian in Liouville space is retrieved via $\kett{\ii \dot{\rho}} = \lindbladian \kett{\rho}$ and reads
\begin{align}
	\lindbladian &= h \sum_j (\sigma^z_j \otimes \identity - \identity \otimes \sigma^z_j) \nonumber\\
	&\hphantom{={}}- J \sum_{\braket{i,j}} (\sigma^x_i \sigma^x_j \otimes \identity - \identity \otimes \sigma^x_i \sigma^x_j) \label{eq:Lindbladian}\\
	&\hphantom{={}}+ \frac{\ii \Gamma}{2} \sum_j (2 \sigma^-_j \otimes \sigma^-_j - \sigma^+_j \sigma^-_j \otimes \identity - \identity \otimes \sigma^+_j \sigma^-_j). \nonumber
\end{align}
This model is illustrated in Fig.~\hyperref[fig:lindbladian]{\ref*{fig:lindbladian}(b)}, with the blue sites ($+$) [red sites ($-$)] corresponding to the Hilbert space on the left [right] side of the tensor product.

In the following, we use \pcst to investigate the limit $|J|\ll |h|, |\Gamma|$. The two terms in the unperturbed part with prefactors $h$ and $\Gamma$ commute, so we can choose
\begin{equation}
\begin{split}
	Q\num{1} &= \frac{1}{2} \sum_j (\sigma^z_j \otimes \identity - \identity \otimes \sigma^z_j)\,,\\
	Q\num{2} &= \sum_{j} (\sigma^+_j \sigma^-_j \otimes \identity + \identity \otimes \sigma^+_j \sigma^-_j - 2 \sigma^-_j \otimes \sigma^-_j)\,,
\end{split}
\label{eq:lindbladnumberoperators}
\end{equation}
with $\epsilon\num{1} = 2 h$ and $\epsilon\num{2} = - \ii \Gamma/2$ and write the perturbation operators $T_\mAlphas$ with two indices.
Further separation of $Q$ into more summands is not possible, because $Q\num{1}$ is the only operator commuting with $\sigma^-_j \otimes \sigma^-_j$.
The perturbation -- decomposed into eigenbasis operators $T_\mAlphas$, as discussed in Sec.~\ref{sec:multipleqptypes}. -- is listed in Tab.~\ref{tab:Lindbladian}.

Even though the perturbation is Hermitian, we have that $T_{(m_1, m_2)}^\dagger \neq T^{\hphantom{\dagger}}_{(-m_1, -m_2)}$, in contrast to the models before.

We can work in a common local bi-orthonormal eigenbasis for both $Q\num{1}$ and $Q\num{2}$, labeled by their respective eigenvalues $q\num{1}$, $q\num{2}$ on site $j$.
The right eigenvectors $\ket{q\num{1}, q\num{2}}$ and the corresponding left eigenvectors $\bra{q\num{1}, q\num{2}}^\textrm{L}$ are given by
\begin{equation}
	\begin{aligned}
	\ket{0, 0} &:= \kett{\downarrow \downarrow},&\ket{- 1, 1} &:= \kett{\downarrow \uparrow},\\  
	\bra{0, 0}^\textrm{L} &:= \bbra{\downarrow \downarrow} + \bbra{\uparrow \uparrow},&\bra{- 1, 1}^\textrm{L} &:= \bbra{\downarrow \uparrow},\\
	\ket{0, 2} &:= \kett{\uparrow \uparrow} - \kett{\downarrow \downarrow},& \ket{+ 1, 1} &:= \kett{\uparrow \downarrow},\\
	\bra{0, 2}^\textrm{L} &:= \bbra{\uparrow \uparrow}, & \bra{+ 1, 1}^\textrm{L} &:= \bbra{\uparrow \downarrow}.
	\end{aligned}
	\label{eq:Lindbladian_eigenbasis}
\end{equation}
Without perturbation, the vector $\ket{\mathrm{ss}} := \otimes_j \ket{0, 0}$ given by $\ket{0, 0}$ on each rung is the stationary state of the Lindbladian and, at the same time, the only eigenvector of $Q\num{1}$ and $Q\num{2}$ corresponding to eigenvalue $0$.
Thus, we consider it as the reference state and introduce hardcore-bosonic annihilation and creation operators
\begin{equation}
	\begin{aligned}
	\anniboth{j} &:= \ketbra{0, 0}{0, 2}^\textrm{L}_j, & \annilr{j} &:= \ketbra{0, 0}{\pm 1, 1}^\textrm{L}_j,\\
	\creaboth{j} &:= \ketbra{0, 2}{0, 0}^\textrm{L}_j, & \crealr{j} &:= \ketbra{\pm 1, 1}{0, 0}^\textrm{L}_j.
	\end{aligned}
\end{equation}
Note that $\ddagger$ does not indicate adjunction since some left eigenvectors are not adjoint to their respective right eigenvectors [see e.g.\ $\ket{0, 0}$ and $\bra{0, 0}^\textrm{L}$ in Eq.~\eqref{eq:Lindbladian_eigenbasis}].
Now we can rewrite the counting operators \eqref{eq:lindbladnumberoperators} in terms of those hardcore bosons as
\begin{equation}
\begin{split}
	Q\num{1} &= \sum_j (\crealeft{j} \annileft{j} - \crearight{j} \anniright{j})\,,\\
	Q\num{2} &= \sum_{j} (2 \creaboth{j} \anniboth{j} + \crealeft{j} \annileft{j} + \crearight{j} \anniright{j})\,,
\end{split}
\end{equation}
and the perturbation operators listed in Table~\ref{tab:Lindbladian_eigenbasis}.

We have calculated the effective Lindbladian up to order 6. Here we explicitly give the \pcst transformed Lindbladian up to \nth{2} in normal ordered form
\begin{equation}
	\lindbladian_\textrm{eff}^{(2, J)} = \lindbladian_{\textrm{eff}, +}^{(2, J)} + \lindbladian_{\textrm{eff}, -}^{(2, J)}\,,
\end{equation}
\begin{widetext}
\begin{equation}
	\begin{split}
	\lindbladian_{\textrm{eff}, \pm}^{(2, J)} &= \pm 2 h \sum_j \crealr{j} \annilr{j} - \frac{\ii \Gamma}{2} \sum_{j} (\creaboth{j} \anniboth{j} + \crealr{j} \annilr{j}) \mp J \sum_{\braket{i,j}} \left[ (\crealr{i} + \creaboth{i} \annirl{i}) (\annilr{j} + \crearl{j} \anniboth{j}) + (i \leftrightarrow j) \right]\\
	&\hphantom{={}}+ \left( \mp \frac{J^2}{2h} \pm \frac{16 h J^2}{16 h^2 - \Gamma^2} - \frac{4 \ii \Gamma J^2}{16 h^2 - \Gamma^2} \right) \sum_{j} \crealr{j} \annilr{j} + \left( \pm \frac{J^2}{h} \mp \frac{16 h J^2}{16 h^2 - \Gamma^2} \right) \sum_{\braket{i,j}} \crealr{i} \annilr{i} \crealr{j} \annilr{j}\\
	&\hphantom{={}}+ \frac{4 \ii \Gamma J^2}{16 h^2 - \Gamma^2} \sum_{\braket{i,j}} \left[ \creaboth{i} \anniboth{i} \crealr{j} \annilr{j} + (i \leftrightarrow j) \right] + \frac{2 \ii \Gamma J^2}{16 h^2 - \Gamma^2} \sum_{\braket{i,j}} (\crealeft{i} \annileft{i} + \crearight{i} \anniright{i}) (\crealeft{j} \annileft{j} + \crearight{j} \anniright{j})\\
	&\hphantom{={}}\pm \frac{J^2}{4h} \sum_{\bbrakett{i,j,k}} \left[ (\annirl{i} \mp \crealr{i} \anniboth{i}) (\crearl{k} \pm \creaboth{k} \annilr{k}) + (i \leftrightarrow k) \right] (\identity - 2 \crealeft{j} \annileft{j} - 2 \crearight{j} \anniright{j})\\
	&\hphantom{={}}\pm \frac{2 J^2}{4 h \mp \ii \Gamma} \sum_{\bbrakett{i,j,k}} \left[ (\crealr{i} + \creaboth{i} \annirl{i}) \annilr{k} + (i \leftrightarrow k) \right] (\creaboth{j} \anniboth{j} - \crearl{j} \annirl{j})\\
	&\hphantom{={}}\mp \frac{2 J^2}{4 h \mp \ii \Gamma} \sum_{\bbrakett{i,j,k}} \left[ (\crealr{i} + \creaboth{i} \annirl{i}) \crearl{k} \anniboth{k} + (i \leftrightarrow k) \right] (\identity - 2 \crealr{j} \annilr{j} - \crearl{j} \annirl{j} - \creaboth{j} \anniboth{j}),
\end{split}
\end{equation}
\end{widetext}
where $\bbrakett{i,j,k}$ indicates three neighboring sites with $j$ being the middle one.
Because the effective Lindbladian is block diagonal with respect to the eigenvalues of both $Q\num{1}$ and $Q\num{2}$, we can discuss the different subspaces separately.
For this example, we select the subspaces with the smallest imaginary part, since the respective density matrices decay slowest and are thus most relevant on large time scales.
This corresponds to low QP numbers of $Q\num{2}$, as it includes decay processes.

{\makegapedcells\begin{table}[t]
	\begin{tabularx}{\columnwidth}{c|c|l}
		$m\num{1}$ & $m\num{2}$ & Terms in $T_\mAlphas$\\ \hline\hline
		$\pm 2$ & $- 2$ & $\mp 2 \crealr{i} \anniboth{i} \annirl{j} \mp 2 \annirl{i} \crealr{j} \anniboth{j}$\\ \hline
		$\pm 2$ & $+ 2$ & $\pm (\crealr{i} + \creaboth{i} \annirl{i}) (\crealr{j} + \creaboth{j} \annirl{j})$\\ \hline
		$0$ & $0$ & \makecell{$(\crealeft{i} + \creaboth{i} \anniright{i}) (\annileft{j} + \crearight{j} \anniboth{j})$\\
		$+ (\annileft{i} + \crearight{i} \anniboth{i}) (\crealeft{j} + \creaboth{j} \anniright{j})$\\
		$- (\crearight{i} + \creaboth{i} \annileft{i}) (\anniright{j} + \crealeft{j} \anniboth{j})$\\
		$- (\anniright{i} + \crealeft{i} \anniboth{i}) (\crearight{j} + \creaboth{j} \annileft{j})$}\\ \hline
		$0$ & $- 2$ & \makecell{$2 \crearight{i} \anniboth{i} \anniright{j} + 2 \anniright{i} \crearight{j} \anniboth{j}$\\
		$- 2 \crealeft{i} \anniboth{i} \annileft{j} - 2 \annileft{i} \crealeft{j} \anniboth{j}$}\\ \hline
		$\pm 2$ & $0$ & \makecell{$\pm (\annirl{i} - \crealr{i} \anniboth{i}) (\crealr{j} + \creaboth{j} \annirl{j})$\\
		$\pm (\crealr{i} + \creaboth{i} \annirl{i}) (\annirl{j} - \crealr{j} \anniboth{j})$}
	\end{tabularx}
	\caption{Perturbation operators of the Lindbladian \eqref{eq:Lindbladian} expressed in the eigenbasis from Eq.~\eqref{eq:Lindbladian_eigenbasis}. These operators are defined via combinations of certain terms, such that the total $\mAlphas$ is correct, i.e., a change of $\mAlphas = (\pm 1, +1)$ can be achieved via $\crealr{j}$ or $\creaboth{j} \annirl{j}$ and a change of $\mAlphas = (\pm 1, -1)$ via $\annirl{j}$ or $\crealr{j} \anniboth{j}$. \label{tab:Lindbladian_eigenbasis}}
\end{table}}

The only stationary state of $\lindbladian$ is $\ket{\mathrm{ss}}$, corresponding to eigenvalue $0$, as all other eigenstates have a non-vanishing eigenvalue.
Therefore, the stationary subspace corresponds to a $1 \times 1$ block of $\lindbladian_\textrm{eff}$.
Without further calculations, we see that
\begin{equation}
	\lindbladian_{(0, 0)}^{(2, J)} = 0\,.
\end{equation}
This actually holds for arbitrary orders, because \mbox{$\bra{\mathrm{ss}}^\textrm{L} T_\m \ket{\mathrm{ss}} = 0$} for all $\m$, where \mbox{$\bra{\mathrm{ss}}^\textrm{L} := \otimes_j \bra{0, 0}^\textrm{L}$}.
Interestingly, this left eigenvector is given by the vectorized identity operator \mbox{$\bbra{\identity} = \sum_{\ket{\varphi}} \bbra{\varphi, \varphi}$} and acting with it is equivalent to performing the trace.
This nicely combines the facts that the left and right eigenvalues form a bi-orthonormal basis and that the stationary state has trace $1$ and is therefore a valid density operator.
This also implies that all other eigenvectors of the Lindbladian have trace $0$ and are not valid density operators.
However, they can be combined with the stationary state to form other density operators.

The two 1QP subspaces correspond to the eigenvalue with the lowest non-zero imaginary part, i.e.,  $(q\num{1}, q\num{2}) = (\pm 1, 1)$.
The effective Lindbladians on this subspace read
\begin{equation}
\begin{split}
	\lindbladian_{(\pm 1, 1)}^{(2, J)} \ket{1_{j, \pm}} &= \pm \left( 2 h \mp \frac{\ii \Gamma}{2} - \frac{J^2}{2 h} + \frac{4 J^2}{4 h \pm \ii \Gamma} \right) \ket{1_{j, \pm}}\\
	&\hphantom{={}} \mp J (\ket{1_{j-1, \pm}} + \ket{1_{j+1, \pm}})\\
	&\hphantom{={}} \mp \frac{J^2}{4 h} (\ket{1_{j-2, \pm}} + \ket{1_{j+2, \pm}}).
\end{split}
\end{equation}

The blocks for higher QP numbers corresponding to eigenvalues with higher imaginary parts -- thus decaying faster -- contain mixtures of different creation and annihilation operators, whose discussion is omitted here.

With this example we have proved that \pcst is able to handle dissipative systems in the form of Lindbladians.
This is done by purification of the Lindblad equation and its density matrices and expressing the dissipative part and a share of the Hamiltonian part as a sum of ladder spectra.
The second-order results show the possibilities of discussing dissipative dynamics in a perturbative fashion, focussing on slowly decaying channels, phrased as QP excitations.

\section{Conclusion}
\label{sec:conclusion}
In this work we have introduced \pcst, which generalizes the established pCUT method to treat quantum many-body systems by high-order linked-cluster expansions in two directions.
Firstly, it is now possible to treat multiple QP types so that one is not restricted to an unperturbed Hamiltonian with an equidistant spectrum corresponding to a single QP counting operator.
Secondly, we have extended the field of application to open quantum systems described by non-Hermitian Hamiltonians or Lindblad operators.
As for the conventional pCUT method, this allows the derivation of model-independent effective Hamiltonians and Lindbladians that commute with the unperturbed part given by arbitrary complex-valued superimposed ladder spectra.
The effective operators are therefore block diagonal and significantly easier to treat.
A linked-cluster expansion in the thermodynamic limit is therefore still possible by design.
This includes in particular white-graph expansions \cite{Coester2015}.
Because the linked-cluster expansion can rely on the same programs for both pCUT as \pcst, we expect application to a wide range of models, including long-ranged or higher-dimensional models, as straightforward.
The potential limitation due to divergences arising from small unperturbed energy differences, was overcome with a tunable generator that does not decouple these subspaces.

On the computational-side, a new challenge (in contrast to the original pCUT) are the additional degrees of freedom introduced by arbitrary ratios of the ladder-spectrum spacings, resulting in a need for analytical expressions of the coefficients.
Although we calculated coefficients in high orders, the \pcst could be further optimized from deeper insights into the actual shape of the coefficients to speed up analytical calculations.

We expect that \pcst will be useful in various directions for future investigations tackling correlated closed and open quantum systems.
A naturally relevant field will be cooperative quantum phenomena involving light and matter degrees of freedom where multiple QP types as well as dissipation and driving are naturally present.

\section*{Acknowledgments}
We acknowledge support by the Deutsche Forschungsgemeinschaft (DFG, German Research Foundation) -- Project-ID 429529648 -- TRR 306 QuCoLiMa (``Quantum Cooperativity of Light and Matter''). KPS acknowledges the support by the Munich Quantum Valley, which is supported by the Bavarian state government with funds from the Hightech Agenda Bayern Plus.

\begin{center}
	Supplementary data~\cite{Data} as well as the source code~\cite{Program} to calculate the \pcst are available online.
\end{center}

\FloatBarrier

\appendix*
\section{Proof of the transformation}
\label{sec:proof_trafo}

In this appendix we prove that the \pcst is convergent, that the effective operator $\operator_\textrm{eff}$ [see Eq.~\eqref{eq:Meff}] is block diagonal and that for Hermitian operators $\operator$ the flowing operator $\operator(\ell)$ is band diagonal.
Thereby, we follow closely the proofs of pCUT \cite{KnetterUhrig2000}.

\paragraph*{Convergence}
Convergence is a necessary property, because it makes the \pcst well defined.
It is given when the coefficients $C_\m = \lim_{\ell \to \infty} F(\ell; \m)$ are unique and finite.
The following proof is only performed for $M(\m) = 0$, while $M(\m) \neq 0$ will be discussed in the next paragraph.
It suffices to consider the limit for the functions $f(\ell; \m) = e^{|M(\m)| \ell} F(\ell; \m)$, since both coincide for $M(\m) = 0$.

Let $\m$ be arbitrary with ${M(\m) = 0}$ and ${(\m_1, \m_2) = \m}$ be an arbitrary breakup.
By using \mbox{$M(\m_1) + M(\m_2) = M(\m) = 0$} we conclude that $M(\m_1) = - M(\m_2)$.
Using that the coefficient function is given by
\begin{align}
	\begin{split}
		f(\ell; \m) &= \sum_{(\m_1, \m_2) = \m} \int e^{- 2 |M(\m_1)| \ell}\\
		&\hphantom{={}} \times 2 \sgn(M(\m_1)) f(\ell; \m_1) f(\ell; \m_2) \dd \ell\,,
	\end{split}
\end{align}
every term in this sum is either vanishing for \mbox{$M(\m_1) = 0$} or contains an exponentially decaying term for ${M(\m_1) \neq 0}$.
Thus the only non-zero contribution to $C_\m$ is the constant of integration.

\paragraph*{Block diagonality}
That the effective operator $\operator_\textrm{eff}$ is block diagonal, i.e, fulfilling $[Q, \operator_\textrm{eff}] = 0$, is an essential aspect of the \pcst.
Because of Eqs.~\eqref{eq:commutator_Q_Tm},~\eqref{eq:Meff} we prove this aspect by showing that $M(\m) \neq 0$ implies $C_\m = 0$.
It suffices to prove by induction that the functions $f$ are of the form
\begin{align}
	\sum_{\mu \geq 0} P_\mu(\ell; \m) e^{-\mu \ell}
	\label{eq:form_function_f}
\end{align}
with non-negative integers $\mu$ and polynomials $P_\mu(\ell; \m)$.
This then implies by definition that the functions $F(\ell; \m) = e^{-|M(\m)| \ell} f(\ell; \m)$ are exponentially decaying for $M(\m) \neq 0$.

The first-order results are given by ${f(\ell; (m_1)) = 1}$ and are thus of the form ~\eqref{eq:form_function_f}.
As induction hypothesis we assume that the statement holds for all $\m$ with \mbox{$|\m| < k$}.
Now let $\m$ be arbitrary with $|\m| = k$.
Because of Eq.~\eqref{eq:RecursiveDifferentialEquation}, the coefficient function is given by
\begin{align}
    \begin{split}
        f(\ell; \m) = \sum_{(\m_1, \m_2) = \m} \int e^{(|M(\m)| - |M(\m_1)| - |M(\m_2)|) \ell}\\
        \times \left[ \sgn(M(\m_1)) - \sgn(M(\m_2)) \right] f(\ell; \m_1) f(\ell; \m_2) \dd \ell\,.
    \end{split}
\end{align}
The term $|M(\m)| - |M(\m_1)| - |M(\m_2)|$ is always non-positive and thus, together with the induction hypothesis, the integrand is of the form discussed above.
Together with the starting condition, the coefficient function is of the form
\begin{align}
	f(\ell; \m) = P_0(\ell; \m) + \sum_{\mu \geq 1} P_\mu(\ell; \m) e^{-\mu \ell}\,.
\end{align}

\paragraph*{Band diagonality}
To say that the flowing operator $\operator(\ell)$ is band diagonal during the flow means that $|M(\m)| > \epsilon_{\textrm{max}}$ implies $F(\ell; \m) = 0$.
This threshold value $\epsilon_{\textrm{max}}$ must be independent on $\ell$ or the order.
Having this property fulfilled is not necessary but computationally desired, because it reduces the number of non-zero coefficient functions.
Here, we prove it for Hermitian operators, where the threshold value $\epsilon_{\textrm{max}}$ is the largest element of $\mathcal{E}$.
Band diagonality does not hold when using the complex signum function~\eqref{eq:complex_sgn}; we will not show that here.

It suffices to prove by induction that $M(\m) > \epsilon_{\textrm{max}}$ implies that $f(\ell; \m) = 0$, which implies per definition that $F(\ell; \m) = 0$.
The \nth{1} order contributions are trivially band diagonal, which is our induction base case.
As induction hypothesis we assume that the statement holds for all $\m$ with $|\m| < k$.
Now let $\m$ be arbitrary with $|\m| = k$ and \mbox{$M(\m) > \epsilon_{\textrm{max}}$} positive.
According to our induction hypothesis, for Eq.~\eqref{eq:RecursiveDifferentialEquation} only terms with $|M(\m_1)|, |M(\m_2)| \leq \epsilon_{\textrm{max}}$ contribute, as $|\m_1|$, $|\m_2| < k$.
Thus we know that $M(\m_1)$ and $M(\m_2)$ are positive due to \mbox{$\epsilon_{\textrm{max}} < M(\m) = M(\m_1) + M(\m_2)$}.
Therefore the term does not contribute at all because of the prefactor containing $\sgn(M(\m_1)) - \sgn(M(\m_2)) = 0$.
We conclude with Eq.~\eqref{eq:RecursiveDifferentialEquation} that $\partial_\ell f(\ell; \m) = 0$ and with the starting conditions also $f(\ell; \m) = 0$ and $F(\ell; \m) = 0$.
The same holds analogously for arbitrary $\m$ with $|\m| = k$ and \mbox{$M(\m) < - \epsilon_{\textrm{max}}$} negative.

\paragraph*{Broad-step signum function}
The shown proofs were made for the generator defined with the generalized sign function in Eq.~\eqref{eq:complex_sgn}.
For the broad-step signum function $\sgn_D$, introduced in Sec.~\ref{sec:broad_signum}, the proofs can be performed in an analogous way.
It turns out that convergence is still preserved, while the block diagonality is no longer fulfilled, as $[Q, \mathcal M_\mathrm{eff}]\neq 0$ in general.
Instead, we can prove the softened property that $C_\m = 0$ for $|M(\m)|>D$, again showing decoupling of subspaces with sufficient distant eigenvalues.
Lastly, the band-diagonality is not fulfilled, again increasing the needed number of calculations, analogous to the non-Hermitian case.

\FloatBarrier 
\newpage
\bibliography{bibliography.bib}

\end{document}